\documentclass[12pt]{iopart}
\usepackage{iopams} 
\usepackage{graphicx, caption}
\usepackage{lipsum} 
\usepackage{setspace}

\begin{document}
\title[Eigenvalue approach for SC-HDM]{An eigenvalue approach to quantum plasmonics based on a self-consistent hydrodynamics method}

\author{Kun Ding and C. T. Chan}

\address{Department of Physics and Institute for Advanced Study, The Hong Kong University of Science and Technology, Clear Water Bay, Hong Kong}
\ead{phchan@ust.hk}

\begin{abstract}
Plasmonics has attracted much attention not only because it has useful properties such as strong field enhancement, but also because it reveals the quantum nature of matter. To handle quantum plasmonics effects, ab initio packages or empirical Feibelman d-parameters have been used to explore the quantum correction of plasmonic resonances. However, most of these methods are formulated within the quasi-static framework. The self-consistent hydrodynamics model offers a reliable approach to study quantum plasmonics because it can incorporate the quantum effect of the electron gas into classical electrodynamics in a consistent manner. Instead of the standard scattering method, we formulate the self-consistent hydrodynamics method as an eigenvalue problem to study quantum plasmonics with electrons and photons treated on the same footing. We find that the eigenvalue approach must involve a global operator, which originates from the energy functional of the electron gas. This manifests the intrinsic nonlocality of the response of quantum plasmonic resonances. Our model gives the analytical forms of quantum corrections to plasmonic modes, incorporating quantum electron spill-out effects and electrodynamical retardation. We apply our method to study the quantum surface plasmon polariton for a single flat interface.
\end{abstract}

\vspace{2pc}
\noindent{\it Keywords}: Eigenvalue approach, Quantum plasmonics, Self-consistent hydrodynamics method, Surface plasmon-polariton dispersion

\maketitle
%
%
\section{Introduction}\label{se:in}

Plasmonic resonances are the intrinsic modes of metallic systems, and these collective excitations have found widespread applications in sensing \cite{anker_biosensing_2008,li_plasmon-enhanced_2014,wong_surface_2014,liang_photonic-plasmonic_2017,zhang_substrate-induced_2011}, photonics \cite{nature_502_80,nm_9_193,ebbesen_extraordinary_1998,lezec_negative_2007,liao_lightning_1982,pendry_negative_2000,fang_subdiffraction-limited_2005}, the lightning rod effect \cite{liao_lightning_1982} and the realization of the perfect lens \cite{pendry_negative_2000,fang_subdiffraction-limited_2005}. Recently, with the development of sophisticated and controllable fabrication techniques for metallic nanoparticles, the quantum nature of electrons has become more obvious and the classical description of local electrodynamics has become inadequate \cite{ciraci_probing_2012,scholl_quantum_2012,teperik_robust_2013,stella_performance_2013,wiener_nonlocal_2012,raza_unusual_2011,dominguez_transformation-optics_2012,luo_surface_2013,marinica_quantum_2012,jin_quantum-spillover_2015,zhu_quantum_2016,esteban_bridging_2012}. Quantum effects, such as the nonlocal response \cite{ciraci_probing_2012,scholl_quantum_2012,teperik_robust_2013,stella_performance_2013,wiener_nonlocal_2012,raza_unusual_2011,dominguez_transformation-optics_2012,luo_surface_2013} and electron spill-out \cite{marinica_quantum_2012,jin_quantum-spillover_2015,zhu_quantum_2016,esteban_bridging_2012}, must be taken into account. The most popular ab initio numerical approach to studying the plasmonic quantum effect is time-dependent density functional theory (TD-DFT) \cite{PhysRevLett_115_137403,PhysRevB_90_161407,PhysRevB_88_155437,PhysRevLett_98_216602,marinica_active_2015,wang_plasmon_2012}. In addition, the hydrodynamics method has also been employed to study quantum plasmonics \cite{ciraci_hydrodynamic_2013,li_electronic_2015,esteban_classical_2015,ford_electromagnetic_1984,PhysRevA.91.053836,toscano_modified_2012,toscano_nonlocal_2013,sobhani_pronounced_2015,varas_quantum_2016,trugler_plasmonics_2017,moradi_plasmon_2015}. While TD-DFT is very useful within the quasi-static framework, the method becomes very computationally demanding if it is coupled with the full Maxwell equations to include electromagnetic (EM) wave characteristics such as retardation \cite{PhysRevB_85_045134,PhysRevB_89_064304}. It is also difficult to extract the physics from fully ab initio numerical simulations, as the sheer volume of information produced can be overwhelming. In order to understand the underlying physics, semi-empirical approaches such as the Feibelman d-parameters \cite{feibelman_surface_1982} have been introduced to model the electron spill-out effect for a single flat interface, and the predicted quantum corrections to surface plasmons have been verified experimentally \cite{Liebsch_metal,PhysRevLett.63.2256,PhysRevLett.64.44,tsuei_normal_1991,sprunger_normal_1992,zacharias_dispersion_1976}. The d-parameter framework has subsequently been extended to more complicated shapes through the boundary element method (BEM) \cite{christensen_quantum_2017}. However, the method of Feibelman d-parameters is still formulated within the quasi-static approximation, meaning that no retardation is incorporated. From an optics and photonics point of view, the Hamiltonian (eigenvalue) approach is used less frequently than scattering formalisms, for the reason that many optics problems are formulated with open boundary conditions. But the eigenvalue approach has proved indispensable in situations where periodic boundary conditions are imposed \cite{PhysRevLett_104_087401,sakoda_phc}, and it is also useful in describing novel properties that are defined using eigenfunctions (such as topological invariants) \cite{lu_topological_2014}. Up to now, essentially all classical Hamiltonian approaches assume that the system must have well-defined macroscopic permittivity and permeability \cite{sakoda_phc}. How to include the quantum effects of electron gases in the Hamiltonian is still an open question.    

Plasmon-polariton excitations, as bound states outside the light cone in a phase space, can be fully described using an eigenvalue formulation. In this article, we attempt to establish the eigenvalue approach based on the self-consistent hydrodynamics model (SC-HDM) to investigate the quantum correction of plasmonic modes. We find that a global operator must exist in the eigenvalue approach in order to make it Hermitian (in the limit of no dissipation). The existence of this global operator means that the response of the system is nonlocal, which is consistent with our physical understanding that classical local electrodynamics is an approximation that will break down when quantum effects kick in. By employing first-order perturbation theory, we give the analytical form of quantum corrections to the plasmonic modes. As an example, we apply our method to study the surface plasmon polaritons (SPPs) of a single flat surface and find that our method agrees with that of the Feibelman d-parameters in the intermediate region of wavenumber space. The paper is organized as follows. In \sref{se:me}, we formulate an eigenvalue approach based on SC-HDM. The related issues on the global operator and the inner product are discussed in \sref{sse:nlo} and \sref{sse:inp}, respectively. The quantum corrections to a general plasmonic mode and the dispersion of quantum SPPs are given in \sref{se:rd}. Conclusions are drawn in \sref{se:cc}.
%
%
\section{Methods}\label{se:me}

To formulate the eigenvalue approach, we employ SC-HDM because it treats electrons and photons on the same footing \cite{PhysRevB_49_8147,PhysRevB_91_115416,toscano_resonance_2015,PhysRevB_93_205405,PhysRevB_95_245434,own_prb}. However, SC-HDM uses the electron density and not the electronic wavefunction as the basic variable, it is not as accurate as local density functionals. Yet previous studies have shown that SC-HDM gives reasonable results compared with ab initio calculations \cite{PhysRevB_49_8147,PhysRevB_91_115416,toscano_resonance_2015,PhysRevB_93_205405,PhysRevB_95_245434,own_prb} but at a much lower computational cost. In the following, we will show the procedure for establishing the eigenvalue approach based on SC-HDM.  

\subsection{Eigenvalue approach}\label{sse:fh}

The first step of SC-HDM involves determining the electronic ground state density that minimizes a density functional $G[n]$ subject to constraints (chemical potential and electron number). This step requires the numerical calculation of the equilibrium electron density $n_0$ and the effective single electron potential $V_{\rm{eff}}$ \cite{own_prb}. Once the ground state density is obtained, the excited state calculations can be performed numerically by coupling the linearized equations of motion for the electron gas with Maxwell equations as \cite{own_prb}
\begin{equation}
\label{eq:s21hel}\nabla\times(\nabla\times\bi{E})-\Big(\frac{\omega}{c}\Big)^2\bi{E}-\rmi\omega\mu_0\bi{J}=0,
\end{equation}
\begin{equation}
\label{eq:s21con}\nabla\cdot\bi{J}-\rmi\omega\rho=0,
\end{equation}
\begin{equation}
\label{eq:s21lr}\frac{en_0}{m_{\rm{e}}}\nabla\Big(\frac{\delta G}{\delta n}\Big)_{\rm{1}}+\frac{e^2 n_0}{m_{\rm{e}}}\bi{E}-(-\rmi\omega+\gamma)\bi{J}=0,
\end{equation}
where $\omega$ is the angular frequency, $c=1/\sqrt{\varepsilon_0\mu_0}$ is the speed of light in vacuum, $m_{\rm{e}}$ ($-e$) is the mass (charge) of an electron, $\bi{E}$ is the microscopic electric field, $\bi{J}$ is the induced current, $\rho$ is the induced charge density, and $\gamma$ is the loss parameter. The first term in \eref{eq:s21lr} gives the nonlocal and spill-out effect of the electron gas, an important effect at the nano-scale that is absent in the classical electromagnetic approaches. In principle, the density functional $G[n]$ has many contributions, including kinetic energy, exchange-correlation, and Coulomb interactions. The exchange-correlation part is explicitly quantum in nature. All other terms depend on the density, which itself is determined variationally by minimizing $G[n]$ with respect to certain constraints. We can view those physical effects that originate from $G[n]$ as quantum in nature. Nevertheless, some of the terms can be derived through a semi-classical approach. We will discuss the explicit forms of $G[n]$ later. Using \eref{eq:s21con}, we see that the currents $\bi{J}$ can be used to replace $n_1$ as the variable in \eref{eq:s21lr}, so we introduce an operator $\hat{\rm{K}}_{\rm{G}}$ defined as
\begin{equation}
\label{eq:s21kg}\frac{c^2}{-\rmi\omega}\hat{\rm{K}}_{\rm{G}}\bi{J}=\frac{en_0}{m_{\rm{e}}}\nabla\Big(\frac{\delta G}{\delta n}\Big)_{\rm{1}},\quad n_1=-\frac{1}{\rmi\omega e}\nabla\cdot\bi{J}.
\end{equation}

We assume that the dispersion of electrons near the Fermi energy is parabolic, namely 
\begin{equation}
\label{eq:s21ef}{E_{\rm{F}}} = \frac{{{\hbar^2}k_{\rm{F}}^2}}{{2{m_{\rm{e}}}}} = \frac{1}{2}{m_{\rm{e}}}v_{\rm{F}}^2 = \frac{{{\hbar^2}}}{{2{m_{\rm{e}}}}}{(3{\pi ^2}{n_{{\rm{ion}}}})^{2/3}},
\end{equation}
where $v_{\rm{F}}$, $k_{\rm{F}}$, and $n_{\rm{ion}}$ are the Fermi velocity, the Fermi wave vector and the ion density, respectively. The ion density is defined via a dimensionless quantity $r_{\rm{s}}$ as
\begin{equation}
\label{eq:s21rs}{n_{{\rm{ion}}}} = \frac{3}{{4\pi {{({r_{\rm{s}}}{a_{\rm{H}}})}^3}}},
\end{equation}
where $a_{\rm{H}}=0.529\AA$ is the Bohr radius. Furthermore, the corresponding Thomas-Fermi screening wave vector and plasma frequency are both defined through the ion density as \cite{kittel_solid}
\begin{equation}
\label{eq:s21qf}{q_{{\rm{TF}}}^2} = \frac{3{{\omega _{\rm{p}}^2}}}{{{v_{\rm{F}}^2}}},\qquad \omega_{\rm{p}}^2 = \frac{{{e^2}{n_{{\rm{ion}}}}}}{{{m_{\rm{e}}}{\varepsilon_0}}}.
\end{equation}

With \eref{eq:s21kg}-\eref{eq:s21qf}, equation \eref{eq:s21lr} could be written as
\begin{equation}
\label{eq:s21lrs}\rmi\omega^{-1}c^2\hat{\rm{K}}_{\rm{G}}\bi{J}+\varepsilon_0\omega_{\rm{p}}^{2}f_0^2\bi{E}-(-\rmi\omega+\gamma)\bi{J}-\rmi\omega_0^2\omega^{-1}\bi{J}=0,
\end{equation}
where $f_0=\sqrt{n_0/n_{\rm{ion}}}$ is a dimensionless quantity. Note that we have added the term $-\rmi\omega_0^2\omega^{-1}\bi{J}$ which stands for interband resonances \cite{PhysRevLett_104_087401}.

It can be shown that \eref{eq:s21hel} and \eref{eq:s21lrs} could be rewritten as
\begin{equation}
\label{eq:s21ff}\bi{A}\bi{x}+\nu\bi{B}\bi{x}+\nu^2\bi{C}\bi{x}=0,
\end{equation}
where $\bi{x}=(\widetilde{\bi{E}},\widetilde{\bi{J}})^{\rm{T}}$, $\nu=\omega/c$, and
\begin{equation}
\label{eq:s21ej}\widetilde{\bi{E}}=\sqrt{\varepsilon_0}\;\bi{E},\quad\widetilde{\bi{J}}=\frac{1}{\omega_{\rm{p}}f_0\sqrt{\varepsilon_0}}\bi{J},
\end{equation}
\begin{equation}
\label{eq:s21Am}\bi{A}=\left(\begin{array}{cc}
\nabla\times\nabla\times & 0\\
0 & \Big(\hat{\rm{K}}_{\rm{G}}-\frac{\omega_0^2}{c^2}\Big)f_0
\end{array}\right),
\end{equation}
\begin{equation}
\label{eq:s21BCm}\bi{B}=\left(\begin{array}{cc}
0 & -\frac{i\omega_{\rm{p}} f_0}{c}\\
-\frac{i\omega_{\rm{p}} f_0^2}{c} & \frac{i\gamma f_0}{c}
\end{array}\right),\quad\bi{C}=\left(\begin{array}{cc}
-1 & 0\\
0 & f_0
\end{array}\right).
\end{equation}

To transform \eref{eq:s21ff} into a Hermitian eigenvalue problem, we introduce an auxiliary vector $\bi{y}=\nu^{-1}\bi{D}\bi{x}$. Then \eref{eq:s21ff} becomes
\begin{equation}
\label{eq:s21ay}\left(\begin{array}{cc}
-\bi{C}^{-1}\bi{B} & -\bi{C}^{-1}\bi{A}\bi{D}^{-1}\\
\bi{D} & 0
\end{array}\right)\left(\begin{array}{c}
\bi{x}\\
\bi{y}
\end{array}\right)=\nu\left(\begin{array}{c}
\bi{x}\\
\bi{y}
\end{array}\right).
\end{equation}

In the absence of losses ($\gamma=0$), the matrix on the left-hand side of \eref{eq:s21ay} would be a Hermitian matrix, so that
\begin{equation}
\label{eq:s21hc}\bi{D}^{\dagger}\bi{D}=-\bi{C}^{-1}\bi{A}=\left(\begin{array}{cc}
\nabla\times\nabla\times & 0\\
0 & f_0^{-1}\Big(\frac{\omega_0^2}{c^2}-\hat{\rm{K}}_{\rm{G}}\Big)f_0
\end{array}\right).
\end{equation}

Note that in \eref{eq:s21hc} we have assumed the existence of the inverse of $\bi{D}$. If we define the operator $\hat{\rm{O}}_{\rm{J}}$ as
\begin{equation}
\label{eq:s21oj}\hat{\rm{O}}_{\rm{J}}^{*}\hat{\rm{O}}_{\rm{J}}\equiv f_0^{-1}\Big(\frac{\omega_0^2}{c^2}-\hat{\rm{K}}_{\rm{G}}\Big)f_0,
\end{equation}
then the matrix $\bi{D}$ could be written as
\begin{equation}
\label{eq:s21Dm}\bi{D}=\left(\begin{array}{cc}
-i\nabla\times & 0\\
0 & \hat{\rm{O}}_{\rm{J}}
\end{array}\right).
\end{equation}

With the help of $\bi{D}$, equation \eref{eq:s21ay} gives the operator $H$ for a photonic system
\begin{equation}
\label{eq:s21Hm}H=\left(\begin{array}{cccc}
0 & -\frac{i\omega_{\rm{p}} f_0}{c} & i\nabla\times & 0\\
\frac{i\omega_{\rm{p}} f_0}{c} & -\frac{i\gamma}{c} & 0 & \hat{\rm{O}}_{\rm{J}}^{*}\\
-i\nabla\times & 0 & 0 & 0\\
0 & \hat{\rm{O}}_{\rm{J}} & 0 & 0
\end{array}\right).
\end{equation}

If we set $\gamma=0$, then $H^{\dagger}=H$, as it must be if there is no dissipation. Meanwhile the auxiliary quantity $\bi{y}$ is evaluated as
\begin{equation}
\label{eq:s21ym}\bi{y}=\frac{c}{\omega}\left(\begin{array}{cc}
-i\nabla\times & 0\\
0 & \hat{\rm{O}}_{\rm{J}}
\end{array}\right)\left(\begin{array}{c}
\widetilde{\bi{E}}\\
\widetilde{\bi{J}}
\end{array}\right)=\left(\begin{array}{c}
\sqrt{\mu_0}\;\bi{H}\\
\frac{-ic}{\omega_{\rm{p}}\sqrt{\varepsilon_0}}\hat{\rm{O}}_{\rm{J}}(f_0^{-1}\bi{P})
\end{array}\right).
\end{equation}

It is easy to see that the above procedure automatically introduces magnetic fields $\bi{H}$ into the formulations. Let us further define $\widetilde{\bi{H}}=\sqrt{\mu_0}\;\bi{H}_1$ and $\widetilde{\bi{P}}=\frac{\omega_0}{\omega_{\rm{p}} f_0\sqrt{\varepsilon_0}}\bi{P}$. Then $\bi{y}$ is simplified as 
\begin{equation}
\label{eq:s21ye}\bi{y}=\left(\begin{array}{c}
\widetilde{\bi{H}}\\
-ic(\hat{\rm{O}}_{\rm{J}}\omega_0^{-1}\widetilde{\bi{P}})
\end{array}\right).
\end{equation}

We can now introduce the four-component vector: $\bi{z}=\left(\widetilde{\bi{E}},\widetilde{\bi{J}},\widetilde{\bi{H}},-ic(\hat{\rm{O}}_{\rm{J}}\omega_0^{-1}\widetilde{\bi{P}})\right)^{\rm{T}}$. The eigenvalue formulation is then
\begin{equation}
\label{eq:s21heig}H\bi{z}=\frac{\omega}{c}\bi{z}.
\end{equation}

We have now completely derived the eigenvalue approach for a photonic system expressed in terms of microscopic electron densities rather than macroscopic permittivity. In the photonic crystal literature, $H$ in equation \eref{eq:s21heig} is often called the Hamiltonian because it can be transformed to the wavevector space and hence forms a Hilbert space. In the following, we discuss the operator $\hat{\rm{O}}_{\rm{J}}$ and the related definition of the inner product within the current framework.

\subsection{Energy functional and nonlocal operator}\label{sse:nlo}

In the previous section, we showed that the eigenvalue approach for photonic systems based on microscopic electron densities could be formulated using the operator  $\hat{\rm{O}}_{\rm{J}}$, which comes from the internal energy $G[n]$ of the electron gas. The explicit form of $G[n]$ should be specified in order to obtain an expression for $\hat{\rm{O}}_{\rm{J}}$. Throughout this paper, we use the form of $G[n]$ used commonly in the literature \cite{PhysRevB_91_115416,own_prb}. $\hat{\rm{K}}_{\rm{G}}$ can then be obtained as
\begin{equation}
\label{eq:s22kg}\hat{\rm{K}}_{\rm{G}}=\frac{v_{\rm{F}}^2}{c^2}\Big[\nabla-2\bi{g}\Big]\Bigg\{\frac{D_{\rm{e1}}}{3}-\frac{C_{2}}{4}\Big[\nabla^2-2\bi{g}\cdot\nabla\Big]\Bigg\}(\nabla\cdot),
\end{equation}
where $\bi{g}\equiv\nabla f_0/f_0$, and the coefficients are 
\begin{equation}
\label{eq:s22de1}D_{\rm{e1}}=f_0^{4/3}+\frac{3}{2}C_2\nabla\cdot\bi{g}-\frac{1}{2}C_1 f_0^{2/3}+3C_0 f_0^{4/3}-C_3 f_0^2,
\end{equation}
\begin{eqnarray}
\label{eq:s22c0}{C_0}&=&\frac{2}{9}\frac{{0.035{{\rm{X}}_1}}}{{{{(0.6024 + {{\rm{X}}_1}f_0^{2/3})}^2}}}\frac{{q_{{\rm{TF}}}^2}}{{n_{{\rm{ion}}}^{2/3}}},\\
\label{eq:s22c1}{C_1}&=&\frac{8}{9}(0.0588 + \frac{{0.035}}{{0.6024 + {{\rm{X}}_1}f_0^{2/3}}})\frac{{q_{{\rm{TF}}}^2}}{{n_{{\rm{ion}}}^{2/3}}},\\
\label{eq:s22c2}{C_2}&=&\frac{{{\lambda_\omega}}}{{k_{\rm{F}}^2}}\qquad \quad {X_1} = 7.8{a_{\rm{H}}}n_{{\rm{ion}}}^{1/3},\\
\label{eq:s22c3}C_3&=&\frac{2}{9}\frac{0.035\mathrm{X}_1^2}{(0.6024+\mathrm{X}_1 f_0^{2/3})^3}\frac{q_{\rm{TF}}^2}{n_{\rm{ion}}^{2/3}}.
\end{eqnarray}

$\lambda_\omega$ comes from the kinetic energy of the inhomogeneous electron gas, and $\lambda_{\omega}^{-1}m_{\rm{e}}$ describes effective masses. It is worth mentioning that there are different choices of the exchange-correlation energy functional in $G[n]$, but they all give essentially the same result \cite{own_prb}.  

In \eref{eq:s21oj}, the first term represents the interband resonance and the second term the quantum correction. The exact form of $\hat{\rm{O}}_{\rm{J}}$ under the energy functional operator $\hat{\rm{K}}_{\rm{G}}$ is not easy to obtain because it involves fractional derivative operators \cite{herrmann_fractional}. The leading differential order in \eref{eq:s22kg} is the first-order derivative, and hence there must exist $\rmd^{1/2}$ terms in $\hat{\rm{O}}_{\rm{J}}$. Mathematically, the fractional derivative of $h(x)$ to order $p$ is often defined by means of Fourier or Mellin integral transforms \cite{herrmann_fractional}. $\rmd^{p}h(x)$ at a point $x_0$ is a local property only when $p$ is an integer, but if $p$ is not an integer, then $\rmd^{p}h(x)$ at $x_0$ not only depends on the values of $h(x)$ near $x_0$, but also relates to $h(x)$ in the whole domain. This indicates that $\hat{\rm{O}}_{\rm{J}}$ must be a global operator, illustrating that the response of the electron gas is nonlocal, which is consistent with our physical understanding as shown in \eref{eq:s21lr}.           

However, if we assume that the second term in \eref{eq:s21oj} is much smaller than the first term, then the operator $\hat{\rm{O}}_{\rm{J}}$ could be approximately written as
\begin{equation}
\label{eq:s22oja}\hat{\rm{O}}_{\rm{J}}\approx\frac{\rmi\omega_0}{c}\Big[\mathbf{I}-\frac{c^2}{2\omega_0^2}f_0^{-1}\hat{\rm{K}}_{\rm{G}}f_0\Big].
\end{equation}

Using \eref{eq:s22oja}, the fourth component of $\bi{z}$ can be explicitly written as
\begin{eqnarray}
\label{eq:s22zp}\frac{-\rmi c}{\omega_{\rm{p}}\sqrt{\varepsilon_0}}\hat{\rm{O}}_{\rm{J}}(f_0^{-1}\bi{P})=\widetilde{\bi{P}}-\frac{\sqrt{\varepsilon_0}\omega_{\rm{p}}f_0}{2e\omega_0}\nabla\Big(\frac{\delta G}{\delta n}\Big)_{\rm{1}}.
\end{eqnarray}

In classical approximations, $G[n]=0$ ($\hat{\rm{K}}_{\rm{G}}=0$). Then \eref{eq:s22zp} recovers the results in the literature \cite{PhysRevLett_104_087401,borisov_wave_2006}. Therefore in addition to the classical Hamiltonian for a dispersive medium, we should include the modulations of energy functional $G[n]$ under EM waves in the formulation.

\subsection{Inner product}\label{sse:inp}

For a completeness method, we define the inner product as
\begin{eqnarray}
\langle\bi{z}_m|\bi{z}_n\rangle&=&\int\mathrm{d}\bi{r}\Bigg\{\frac{1}{2}\mu_0\bi{H}_m^*\cdot\bi{H}_n+\frac{1}{2}\varepsilon_0\bi{E}_m^*\cdot\bi{E}_n\nonumber\\
\label{eq:s23inp}&&+\frac{1}{2\omega_{\rm{p}}^2\varepsilon_0}\Big[\frac{1}{f_0^2}\bi{J}_m^*\cdot\bi{J}_n+c^2(\hat{\rm{O}}_{\rm{J}}^{*}f_0^{-1}\bi{P}_m^{*})\cdot(\hat{\rm{O}}_{\rm{J}}f_0^{-1}\bi{P}_n)\Big]\Bigg\},
\end{eqnarray}
where the factor of $1/2$ is the normalized factor that makes the product the total energy of a certain state \cite{PhysRevLett_104_087401,sakoda_phc}. Using \eref{eq:s22zp}, the inner product \eref{eq:s23inp} could be explicitly written as
\begin{eqnarray}
\langle\bi{z}_m|\bi{z}_n\rangle&=&\int\mathrm{d}\bi{r}\Bigg\{\frac{1}{2}\mu_0\bi{H}_m^*\cdot\bi{H}_n+\frac{1}{2}\varepsilon_0\bi{E}_m^*\cdot\bi{E}_n\nonumber\\
&&+\frac{1}{2\omega_{\rm{p}}^2f_0^2\varepsilon_0}\bi{J}_m^*\cdot\bi{J}_n+\frac{\omega_0^2}{2\omega_{\rm{p}}^2f_0^2\varepsilon_0}\bi{P}_m^{*}\cdot\bi{P}_n\nonumber\\
\label{eq:s23ain}&&-\frac{1}{4e}\Big[\bi{P}_m^{*}\cdot\nabla\Big(\frac{\delta G}{\delta n}\Big)_{n,1}+\bi{P}_n\cdot\nabla\Big(\frac{\delta G}{\delta n}\Big)^{*}_{m,1}\Big]\Bigg\}.
\end{eqnarray}

The bracket in the integral is the energy density of a particular photonic state (for $m=n$) with the internal energy of the electron gas taken into account. Let $W_n$ denote the bracket. Then the orthogonal condition \eref{eq:s23inp} used within this method is
\begin{equation}
\label{eq:s23ort}\langle\bi{z}_m|\bi{z}_n\rangle=\delta_{mn}\Big(\int\mathrm{d}\bi{r}\,W_n\Big).
\end{equation}

Note once again that our method is an eigenvalue approach, while others are based on the scattering approach \cite{toscano_resonance_2015}. Computationally, the incident EM wave must be specified for the scattering approach, and the focus is on the scattering of the incident wave by the plasmonic object. Our method solves an eigenvalue problem, similar to band dispersion calculations, and the output consists of the eigenmodes (e.g. surface plasmons) supported by the system. Thus the eigenvalue approach shown in this section is quite different from the scattering approach although the partial differential equations are the same.
%
%
\section{Results and discussions}\label{se:rd}

We have established the eigenvalue approach based on SD-HDM. In this section, we give an explicit expression for quantum corrections to certain plasmonic modes using first-order perturbation theory, and discuss the dispersion of quantum SPPs for a single flat interface.

\subsection{Quantum correction of plasmonic modes}\label{sse:qc}

To study the quantum correction of particular plasmonic modes, we can treat the terms involving quantum corrections and loss in \eref{eq:s21Hm} as the perturbation potential, namely $H=H_0+H_{1}$ in which 
\begin{equation}
\label{eq:s31h0}H_0=\left(\begin{array}{cccc}
0 & -\frac{i\omega_{\rm{p}}f_{\rm{ion}}}{c} & i\nabla\times & 0\\
\frac{i\omega_{\rm{p}}f_{\rm{ion}}}{c} & 0 & 0 & -\frac{i\omega_0}{c}\\
-i\nabla\times & 0 & 0 & 0\\
0 & \frac{i\omega_0}{c} & 0 & 0
\end{array}\right),
\end{equation}
\begin{equation}
\label{eq:s31h1}H_1=\left(\begin{array}{cccc}
0 & -\frac{i\omega_{\rm{p}}(f_0-f_{\rm{ion}})}{c} & 0 & 0\\
\frac{i\omega_{\rm{p}}(f_0-f_{\rm{ion}})}{c} & -\frac{i\gamma}{c} & 0 & \frac{ic}{2\omega_0}f_0^{-1}\hat{\rm{K}}_{\rm{G}}f_0\\
0 & 0 & 0 & 0\\
0 & -\frac{ic}{2\omega_0}f_0^{-1}\hat{\rm{K}}_{\rm{G}}f_0 & 0 & 0
\end{array}\right),
\end{equation}
where $f_{\rm{ion}}=\sqrt{n_{\rm{j}}/n_{\rm{ion}}}$, and $n_{\rm{j}}$ describes the ion density in the jellium model \cite{own_prb}. Mathematically, $f_{\rm{ion}}=1_{\rm{M}}(\bi{r})$ is the indicator function and $\rm{M}$ stands for the union of all metallic domains. Suppose the eigenmodes of $H_0$ are $H_0\bi{z}_m^{(0)}=c^{-1}\omega_m^{(0)}\bi{z}_m^{(0)}$ ($m$ is the plasmonic mode index). Then perturbation theory gives the first-order corrections to the plasmon resonance frequency $\omega_m^{(0)}$ as
\begin{eqnarray}
\fl\label{eq:s31w1e1}c^{-1}\omega_m^{(1)}&=&\frac{\langle\bi{z}_m^{(0)}|\bi{H}_1|\bi{z}_m^{(0)}\rangle}{\langle\bi{z}_m^{(0)}|\bi{z}_m^{(0)}\rangle}\\
\fl&=&\frac{1}{\int\mathrm{d}\bi{r}\,W_m^{(0)}}\int\mathrm{d}\bi{r}\;\frac{1}{2}\Bigg\{-\frac{i\gamma}{c}\widetilde{\bi{J}}_{m}^{*}\cdot\widetilde{\bi{J}}_{m}+\frac{i\omega_{p}(f_0-f_{\rm{ion}})}{c}\Big(\widetilde{\bi{J}}_{m}^{*}\cdot\widetilde{\bi{E}}_{m}-\widetilde{\bi{E}}_{m}^{*}\cdot\widetilde{\bi{J}}_{m}\Big)\nonumber\\
\fl\label{eq:s31w1e2}&&\qquad\qquad\qquad\qquad\quad+\frac{ic}{2\omega_0}\Big[\widetilde{\bi{J}}_{m}^{*}f_0^{-1}\hat{\rm{K}}_{\rm{G}}f_0\widetilde{\bi{P}}_{m}-\widetilde{\bi{P}}_{m}^{*}f_0^{-1}\hat{\rm{K}}_{\rm{G}}f_0\widetilde{\bi{J}}_{m}\Big]\Bigg\},
\end{eqnarray}
where $W_m^{(0)}$ denotes the classical energy density of the photonic state, i.e. setting $G[n]=0$ in \eref{eq:s23ain}. For simplicity, we have omitted the superscript $(0)$ in the fields $(\bi{E},\bi{H},\bi{P},\bi{J})$ and will use this notation in the following. Before proceeding, let us discuss the conditions under which the perturbation approximation holds. The condition $|H_1|\ll|H_0|$ indicates that the variations in ground state densities and the surface-to-volume ratio of the nanoparticles should be small. 

In \eref{eq:s31w1e2} there are three terms: the first one is due to loss, and the other two terms are quantum corrections. The first term is a purely imaginary number, which could be evaluated as
\begin{equation}
\label{eq:s31w1ls}\omega_m^{(1)}=\frac{1}{\int\mathrm{d}\bi{r}\,W_m^{(0)}}\;\frac{-i\gamma}{2\omega_p^2\epsilon_0}\int\mathrm{d}\bi{r}\;\frac{1}{f_{\rm{ion}}}\Big|\bi{J}_m\Big|^2.
\end{equation}
This is consistent with classical EM results in \cite{PhysRevLett_104_087401}. In order to relate the other two terms to the Feibelman $d$-parameters ($d_{\perp}$ and $d_{\parallel}$) defined in quasi-static approximations which are used to explain quantum corrections to plasmonic modes \cite{feibelman_surface_1982}, we introduce the quantum correction factor $\mathcal{F}$ as follows:
\begin{equation}
\label{eq:s31qf}\mathcal{F}=4\Omega_0\Omega_1,\qquad\Omega_0=\frac{\omega_m^{(0)}}{\omega_{\rm{p}}},\quad\Omega_1=\frac{\omega_{m}^{(1)}}{\omega_{\rm{p}}}.
\end{equation}

Within the quasi-static limit, BEM gives the quantum correction factor as $\mathcal{F}=d_{\perp}\Lambda_{\perp}+d_{\parallel}\Lambda_{\parallel}$ and 
\begin{eqnarray}
\label{eq:s31dperp}d_{\perp}=\frac{\int x\rho_1\mathrm{d}x}{\int\rho_1\mathrm{d}x}&\qquad&\Lambda_{\perp}=\frac{(\Lambda^{(0)})^2-1}{2\epsilon_0}\frac{\langle\sigma^{(0)}|\sigma^{(0)}\rangle}{\langle\phi^{(0)}|\sigma^{(0)}\rangle},\\
\label{eq:s21dpara}d_{\parallel}=\int(f_0^2-f_{\rm{ion}}^2)\mathrm{d}x&\qquad&\Lambda_{\parallel}=2\epsilon_0\frac{\langle\nabla_{\parallel}\phi^{(0)}|\nabla_{\parallel}\phi^{(0)}\rangle}{\langle\phi^{(0)}|\sigma^{(0)}\rangle},
\end{eqnarray}
\begin{equation}
\label{eq:s31lb0}\Omega_0^2=\frac{1}{2}(1+\Lambda^{(0)}),
\end{equation}
where $\sigma^{(0)}$ and $\phi^{(0)}$ are the surface charge density and the electrostatic potential in the classical model \cite{christensen_quantum_2017}. Within the quasi-static framework, the quantum correction factor $\mathcal{F}$ can be factorized into the shape factor $\Lambda_{\perp,\parallel}$ and the electron spill-out factor $d_{\perp,\parallel}$. However, while the d-parameters are well defined for a single flat interface,  they are not applicable to the description of nanoparticles with complex geometries, especially those with sharp corners.

Next, we give the explicit form of the quantum correction factor within our method. Using the properties in the classical model
\begin{equation}
\label{eq:s31cls}\bi{J}_m=\rmi\omega^{-1}\omega_{\rm{p}}^2\epsilon_0 f_{\rm{ion}}\bi{E}_m,\qquad\bi{P}_m=-\frac{\epsilon_0\omega_{\rm{p}}^{2}f_{\rm{ion}}}{\omega^2}\bi{E}_m,
\end{equation}
we could explicitly obtain
\begin{eqnarray}
\fl\mathcal{F}&=&\frac{4\epsilon_0}{\int\mathrm{d}\bi{r}\,W_m^{(0)}}\;\int\mathrm{d}\bi{r}\;\Bigg\{(f_0-f_{\rm{ion}})\Big|\bi{E}_m\Big|^2\nonumber\\
\fl\label{eq:s31qfp}&&\qquad\qquad-\frac{3c^2/v_{\rm{F}}^2}{4\Omega_0^2 q_{\rm{TF}}^2}\frac{1}{\omega_0 f_0}\Big[\bi{E}_m^{*}\cdot\hat{\rm{K}}_{\rm{G}}(\omega_0 f_0\bi{E}_m)+\omega_0\bi{E}_m^{*}\cdot\hat{\rm{K}}_{\rm{G}}(f_0\bi{E}_m)\Big]\Bigg\}.
\end{eqnarray}

The first term in the integral depends on the electrostatic surface dipole and electric field intensities, while the second term originates from the nonlocal operators, indicating that the second term depends more on details of the mode than the first one. To obtain physical insights into this correction, we keep the leading derivative term in $\hat{\rm{K}}_{\rm{G}}$, namely
\begin{equation}
\label{eq:s31kgld}\hat{\rm{K}}_{\rm{G}}\approx-\frac{2v_{\rm{F}}^2D_{e1}}{3c^2}\bi{g}(\nabla\cdot).
\end{equation} 
Then \eref{eq:s31qfp} becomes
\begin{equation}
\fl\label{eq:s31qfap}\mathcal{F}=\frac{4\epsilon_0}{\int\mathrm{d}\bi{r}\,W_m^{(0)}}\int\mathrm{d}\bi{r}\;\Bigg\{(f_0-f_{\rm{ion}})\Big|\bi{E}_m\Big|^2+\frac{D_{e1}}{\Omega_0^2 q_{\rm{TF}}^2}(\bi{E}_m^{*}\cdot\bi{g})\Big[\nabla\cdot\bi{E}_m+\bi{g}\cdot\bi{E}_m\Big]\Bigg\},
\end{equation}
where we set $\omega_0=0$ because of our focus on plasmonic systems. Firstly, according to our method, the contributions from the electron spill-out effect and shape effects to the quantum corrections are mixed. Moreover, the first term in the bracket does not have any feature length explicitly, but the second term carries the Thomas-Fermi screening length scale $\ell_{\rm{TF}}\equiv q_{\rm{TF}}^{-1}$. For example, the typical value of $\ell_{\rm{TF}}$ is $0.675\AA$ for sodium. Secondly, the major contribution to the integration comes from the surface region of the metallic particles, because $(f_0-f_{\rm{ion}})$ is only nonzero near the surface region, $D_{e1}$ is zero in the bulky vacuum domain, and $\bi{g}$ is zero in the bulky metal domain. 

To seek a clearer understanding, we use the following relations:
\begin{equation}
\label{eq:s31pe}\epsilon_0\nabla\cdot\bi{E}_m=\rho^{(0)}_{P,m}+\sigma^{(0)}_{P,m}\delta_{\rm{S}}(\bi{r}),
\end{equation}
\begin{equation}
\label{eq:s31ge}\Big(\bi{g}\cdot\bi{E}_{m}\Big)(\bi{s})=\Big(\bi{g}\cdot\bi{E}_{m,\parallel}\Big)(\bi{s})-\frac{\Lambda^{(0)}g_{\perp}(\bi{s})}{2\epsilon_0}\sigma^{(0)}_{P,m},\qquad\bi{s}\in\partial\rm{M},
\end{equation} 
where $g_{\perp}=\bi{g}\cdot\hat{\bi{n}}$, $\bi{g}=|\bi{g}|\hat{\bi{g}}$, and $\rho^{(0)}_{P,m}$ and $\sigma^{(0)}_{P,m}$ are bulk and surface charge densities, and $\parallel$/$\perp$ stands for tangential/normal directions of the surface $\partial\rm{M}$. Then we could simplify the quantum correction factor $\mathcal{F}$ in \eref{eq:s31qfap} as
\begin{equation}
\fl\label{eq:s31qf2}\mathcal{F}=\frac{4\epsilon_0}{\int\mathrm{d}\bi{r}\,W_m^{(0)}}\left\{\begin{array}{l}
\int\mathrm{d}\bi{r}\;\Bigg[(f_0-f_{\rm{ion}})\Big|\bi{E}_m\Big|^2\Bigg]+\ell_{\rm{c},m}^2\int\mathrm{d}\bi{r}\;\Bigg[D_{e1}|\bi{g}|^2\Big|\hat{\bi{g}}\cdot\bi{E}_{m}\Big|^2\Bigg]\\
+\frac{\ell_{\rm{c},m}^2}{\epsilon_0}\int_{\partial\rm{M}}\mathrm{d}A(\bi{s})\;\Bigg[D_{e1}|\bi{g}|\Big(\hat{\bi{g}}\cdot\bi{E}_{m,\parallel}^{*}\Big)\sigma^{(0)}_{P,m}\Bigg](\bi{s})\\
-\frac{\Lambda^{(0)}\ell_{\rm{c},m}^2}{2\epsilon_0^2}\int_{\partial\rm{M}}\mathrm{d}A(\bi{s})\;\Bigg[D_{e1}g_{\perp}\Big|\sigma^{(0)}_{P,m}\Big|^2\Bigg](\bi{s})
\end{array}\right\},
\end{equation}
where the feature length $\ell_{\rm{c},m}^2=\ell_{\rm{TF}}^2\Omega_0^{-2}$ is determined from classical results and material properties.

There are four terms in \eref{eq:s31qf2}. The first two are volume integrals in all of the space, but the contributions mainly come from the surface region of the metallic particles. Meanwhile, the last two terms are integrals on the metal surface. Therefore the quantum correction factor $\mathcal{F}$ is nonnegligible only for particles with a large surface-to-volume ratio.

\subsection{Dispersion of quantum surface plasmon polaritons}\label{sse:fs}

\begin{figure}
\centering
\captionsetup{width=\linewidth}
\includegraphics[width=5.5in]{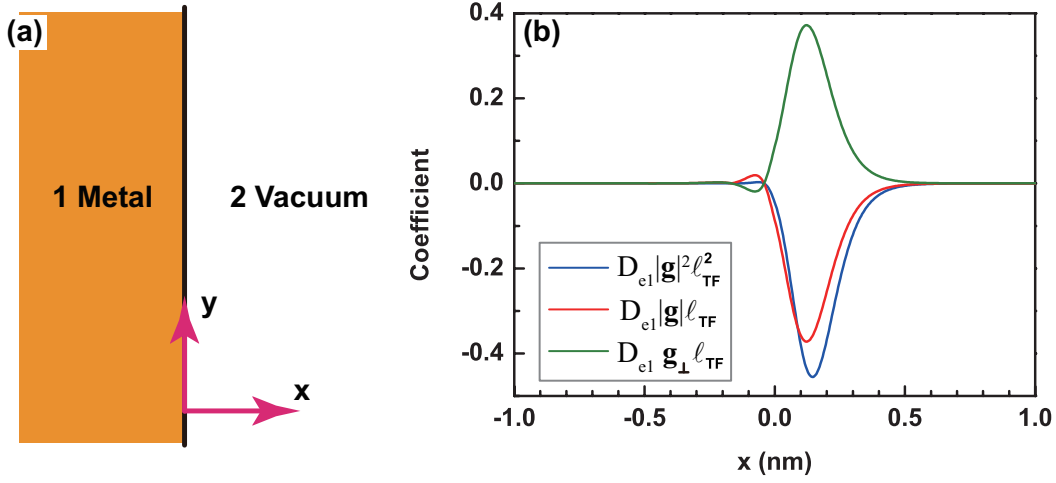}%
\caption{(a) Schematic diagram of a single metallic interface formed at $x=0\rm{nm}$ by metal ($x<0\rm{nm}$) and vacuum ($x>0\rm{nm}$). (b) The values of $D_{e1}|\bi{g}|^2\ell_{\rm{TF}}^2$, $D_{e1}|\bi{g}|\ell_{\rm{TF}}$ and $D_{e1}g_{\perp}\ell_{\rm{TF}}$ as a function of $x$ near the surface region for a single interface shown in (a). The metal used here is sodium with $r_s=4$ and $\lambda_\omega=0.12$.}
\label{fig:s32coef}
\end{figure}

In this section, we aim to apply our method to a single plasmonic interface as shown in \fref{fig:s32coef}(a). To see the coefficients in \eref{eq:s31qf2}, we plot $D_{e1}|\bi{g}|^2\ell_{\rm{TF}}^2$, $D_{e1}|\bi{g}|\ell_{\rm{TF}}$ and $D_{e1}g_{\perp}\ell_{\rm{TF}}$ of this single interface case in \fref{fig:s32coef}(b). It is clear that all of the coefficients are nonzero near the surface as expected. 

The next task is to explicitly write down the classical results of SPPs for the system in \fref{fig:s32coef}(a). The magnetic fields are given as
\begin{eqnarray}
\label{eq:s32hf1}\bi{H}^{1}&=&\hat{\bi{z}}H_0\exp{(\rmi qy+\alpha_1 x-\rmi\omega t)}\qquad q^2-\alpha_1^2=k_0^2\epsilon_r,\\
\label{eq:s32hf2}\bi{H}^{2}&=&\hat{\bi{z}}H_0\exp{(\rmi qy-\alpha_2 x-\rmi\omega t)}\qquad q^2-\alpha_2^2=k_0^2,
\end{eqnarray} 
where $k_0=\omega/c$, $q$ is the parallel wavenumber, and $\alpha_{1,2}>0$. The related electric fields are 
\begin{eqnarray}
\label{eq:s32ef1}\bi{E}^{1}&=&\frac{\rmi}{\omega\epsilon_{0}\epsilon_{r}}\Big[\hat{\bi{x}}\rmi q-\hat{\bi{y}}\alpha_1\Big]H_0\exp{(\rmi qy+\alpha_1 x-\rmi\omega t)},\\
\label{eq:s32ef2}\bi{E}^{2}&=&\frac{\rmi}{\omega\epsilon_{0}}\Big[\hat{\bi{x}}\rmi q+\hat{\bi{y}}\alpha_2\Big]H_0\exp{(\rmi qy-\alpha_2 x-\rmi\omega t)}.
\end{eqnarray}
The boundary condition of $\bi{E}_{\parallel}$ gives the dispersion relation of SPPs as
\begin{equation}
\label{eq:s32dr0}\Omega_0^2=\frac{1}{2}+\eta^2-\sqrt{\frac{1}{4}+\eta^4},
\end{equation}
where the dimensionless quantities $\xi=q/k_0$ and $\eta=\xi\Omega_0=qc/\omega_{p}$ are introduced to simplify the expressions. 

The quantum correction factor $\mathcal{F}$ in \eref{eq:s31qf2} for the single interface case can now be given term by term. The first term in \eref{eq:s31qf2} is
\begin{eqnarray}
\label{eq:s32qf1d}\mathcal{F}_1&=&\frac{4\epsilon_0}{\int\mathrm{d}\bi{r}\,W_m^{(0)}}\int\mathrm{d}\bi{r}\;\Bigg[(f_0-f_{\rm{ion}})\Big|\bi{E}_m\Big|^2\Bigg]\\
\label{eq:s32qf1f}&{=}&\frac{8\alpha_2}{\Theta(\xi^{-1})}\Big(2-\xi^{-2}\Big)\Big[\Omega_0^2(2-\xi^{-2})\mathrm{I}_{1-}+\mathrm{I}_{1+}\Big],
\end{eqnarray}
with the following notations:
\begin{equation}
\label{eq:s32THd}\Theta(\zeta)=-\zeta^3+4\zeta^2-6\zeta+4,
\end{equation}
\begin{equation}
\label{eq:s32I1m}\mathrm{I}_{1-}=\int_{-\infty}^{0}\mathrm{d}x\,(f_0-f_{\rm{ion}})\exp{(2\alpha_1 x)},
\end{equation}
\begin{equation}
\label{eq:s32I1p}\mathrm{I}_{1+}=\int_{0}^{+\infty}\mathrm{d}x\,(f_0-f_{\rm{ion}})\exp{(-2\alpha_2 x)}.
\end{equation}
The second term in \eref{eq:s31qf2} is
\begin{eqnarray}
\label{eq:s32qf2d}\mathcal{F}_2&=&\frac{4\epsilon_0}{\int\mathrm{d}\bi{r}\,W_m^{(0)}}\ell_{\rm{c},m}^2\int\mathrm{d}\bi{r}\;\Bigg[D_{e1}|\bi{g}|^2\Big|\hat{\bi{g}}\cdot\bi{E}_{m}\Big|^2\Bigg]\\
\label{eq:s32qf2f}&{=}&\frac{8\alpha_2}{\Theta(\xi^{-1})}\ell_{\rm{TF}}^2\Big(2-\xi^{-2}\Big)\Big[(1-\xi^{-2})\mathrm{I}_{2-}+(1-\xi^{-2})^{-1}\mathrm{I}_{2+}\Big],
\end{eqnarray}
where
\begin{equation}
\label{eq:s32I2m}\mathrm{I}_{2-}=\int_{-\infty}^{0}\mathrm{d}x\,D_{e1}|\bi{g}|^2\exp{(2\alpha_1 x)},
\end{equation}
\begin{equation}
\label{eq:s32I2p}\mathrm{I}_{2+}=\int_{0}^{+\infty}\mathrm{d}x\,D_{e1}|\bi{g}|^2\exp{(-2\alpha_2 x)}.
\end{equation}

The third term of the quantum correction factor $\mathcal{F}$ in \eref{eq:s31qf2} for the single interface case is zero due to the fact that $\bi{g}$ is perpendicular to the surface. However, for particles have corners, $\bi{g}$ must have a parallel component to the surface, so this part of the quantum correction factor can become prominent. The fourth term of the quantum correction factor $\mathcal{F}$ in \eref{eq:s31qf2} for the single interface case is
\begin{eqnarray}
\label{eq:s32qf4d}\mathcal{F}_4&=&\frac{4\epsilon_0}{\int\mathrm{d}\bi{r}\,W_m^{(0)}}\frac{-\Lambda^{(0)}\ell_{\rm{c},m}^2}{2\epsilon_0^2}\int_{\partial\rm{M}}\mathrm{d}A(\bi{s})\;\Bigg[D_{e1}g_{\perp}\Big|\sigma^{(0)}_{P,m}\Big|^2\Bigg](\bi{s})\\
\label{eq:s32qf4f}&{=}&\frac{4\alpha_2}{\Theta(\xi^{-1})}(-\ell_{\rm{TF}}^2)\frac{(1-2\xi^2)^2}{\xi^4(1-\xi^2)}\Big(D_{e1}g_{\perp}\Big)\Big|_{x=0}.
\end{eqnarray}

\begin{figure}
\centering
\captionsetup{width=\linewidth}
\includegraphics[width=6.0in]{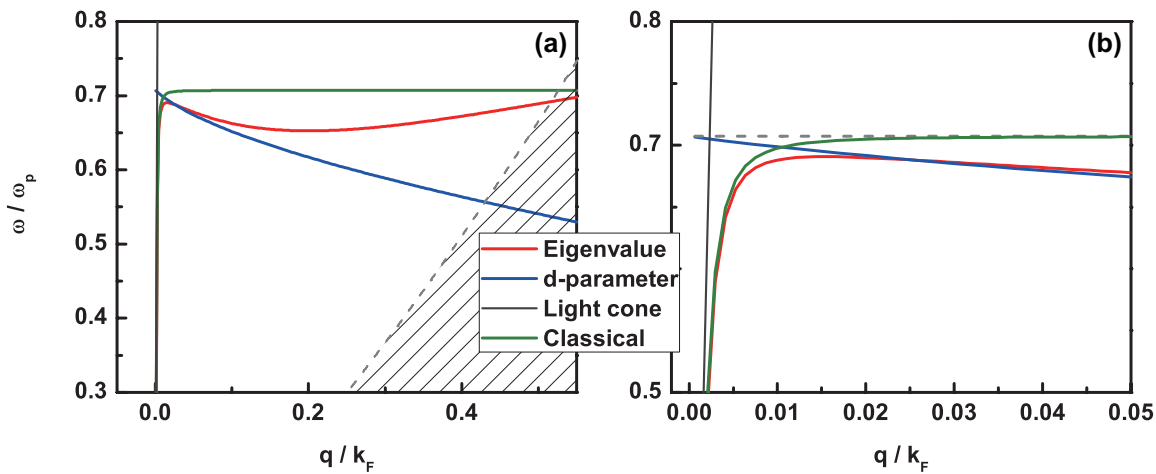}%
\caption{(a) Dispersion relations of surface plasmon polaritons calculated using our model (solid red line), d-parameters (solid blue line), and the classical model (solid green line). The light cone is plotted with the solid dark gray line. The shaded area is the electron-hole pair region, which means that loss is prominent. (b) An enlarged view of the dispersion shown in (a) in the small wavenumber limit.}
\label{fig:s32disp}
\end{figure}

The full numerical results of the SPP dispersion with quantum correction factor $\mathcal{F}$ are shown by the solid red line in \fref{fig:s32disp}(a). For comparison, we also plot the SPP dispersion of the classical model and the d-parameter method using green and blue lines, respectively \cite{feibelman_surface_1982,christensen_quantum_2017,own_prb,Mahan_many}. A magnified version of the SPP dispersion in the small wavenumber limit is shown in \fref{fig:s32disp}(b). We see that the d-parameter approach works well in the intermediate region of $q$ space.

Before ending this section, let us discuss the asymptotic behaviors of $\mathcal{F}$. Firstly, in the limit of a small $q$, namely $\xi\rightarrow 1^{+}$, $\mathcal{F}_{1,2,4}\approx 0$ because $\alpha_2\rightarrow 0$, as shown by the red and green lines in \fref{fig:s32disp}(b). Secondly, in the large $q$ limit, namely $\xi\gg1$,
\begin{equation}
\label{eq:s32qf1lq}\mathcal{F}_1\approx 4q\Big[2\Omega_0^2\mathrm{I}_{1-}+\mathrm{I}_{1+}\Big],
\end{equation}
\begin{equation}
\label{eq:s32qf2lq}\mathcal{F}_2\approx 4q\ell_{\rm{TF}}^2\Big[\mathrm{I}_{2-}+\mathrm{I}_{2+}\Big],
\end{equation}
\begin{equation}
\label{eq:s32qf4lq}\mathcal{F}_4\approx 4q\ell_{\rm{TF}}^2\xi^{-2}\Big(D_{e1}g_{\perp}\Big)\Big|_{x=0}.
\end{equation}

It is not difficult to see that $\mathcal{F}_4\ll\mathcal{F}_{1,2}$, and if we are only concerned with the behavior near the plasmon resonance ($\approx\omega_{p}/\sqrt{2}$), then the quantum correction factor could be further simplified as 
\begin{equation}
\label{eq:s32qftlq}\mathcal{F}=4q\Big[\mathrm{I}_{1}+\ell_{\rm{TF}}^2\mathrm{I}_{2}\Big],
\end{equation}
where $\mathrm{I}_{1}=\mathrm{I}_{1,-}+\mathrm{I}_{1,+}$ and $\mathrm{I}_{2}=\mathrm{I}_{2,-}+\mathrm{I}_{2,+}$. These two integrals are infinite series in $q$, and the leading order in $q$ is a constant, indicating that $\mathcal{F}$ in \eref{eq:s32qftlq} is linear in $q$. This is numerically shown by the red and blue lines in \fref{fig:s32disp}(b).
%
%
\section{Conclusions}\label{se:cc}

We formulated an eigenvalue approach for plasmonic resonances using the self-consistent hydrodynamics model. We showed that the Hamiltonian carries a global operator, indicating that the response of quantum plasmonic resonances is highly nonlocal at the nano-scale. We derived the analytical forms of quantum corrections to a general plasmonic mode. The calculated dispersions of quantum surface plasmon polaritons for a single interface show that in the intermediate $q/k_{\rm{F}}$ region, our results agree well with the Feibelman d-parameter method.  

\ack
We thank Prof. Hai Zhang for helpful discussions. This work is supported by the Research Grants Council of Hong Kong (grant No. AOE/P-02/12).

\section*{References}
\bibliography{ref_eigen}

\end{document}